\documentclass[12pt,leqno]{article}
\usepackage{amssymb,amsthm,amsmath,latexsym,euscript}
\usepackage{fancybox}
\usepackage{longtable}
\usepackage{listings}
\lstdefinelanguage{magma}
{mathescape}
\newtheorem{thm}{Theorem}[section]\normalfont\huge
\newtheorem{prop}[thm]{Proposition}

\newtheorem{cor}[thm]{Corollary}

\newcommand{\pf}{\noindent \textbf{Proof.} }
\renewcommand{\qed}{\hfill $\Box$ \\}

\newtheorem{ex}[thm]{Example}

\numberwithin{equation}{section}

\makeatletter
\renewcommand{\@biblabel}[1]{#1\hfill \hspace{-0.2cm}}
\makeatother

\def\x{{\mathbf x}}
\def\y{{\mathbf y}}

\def\u{{\mathbf u}}
\def\v{{\mathbf v}}
\def\c{{\mathbf c}}

\def\r{{\mathbf r}}
\def\g{{\mathbf g}}

\def\t{{\mathbf{t}}}

\def\0{{\mathbf 0}}
\def\C{{\mathcal C}}

\newcommand{\F}{\mathbb{F}_p}

\newcommand{\Aut}{Aut}

\author{
Altaf Alshuhail\\ 
Department of Mathematics \\
College of Science\\
University of Ha’il\\
Ha’il,Saudi Arabia\\
{\small Email: \texttt{ad.alshuhail@uoh.edu.sa}}\\
\\
Rowena Alma Betty\\
Institute of Mathematics \\
University of the Philippines\\
Diliman, Quezon City\\
Philippines \\
\small{Email: \texttt{rabetty@math.upd.edu.ph}}\\
\\
Lucky Galvez\\
Institute of Mathematics \\
University of the Philippines\\
Diliman, Quezon City\\
Philippines \\
{\small Email: \texttt{legalvez@math.upd.edu.ph}}\\
}

\title{Duality of Codes over Non-unital Rings of Order Six}

\date{ }

\begin{document}

\maketitle


\begin{abstract}
We present some basic theory on the duality of codes over two non-unital rings of order $6$, namely $H_{23}$ and $H_{32}$. For a code $\C$ over these rings, we associate a binary code $\C_a$ and a ternary code $\C_b$. We characterize self-orthogonal, self-dual and quasi self-dual (QSD) codes over these rings using the codes $\C_a$ and $\C_b$. In addition, we present a building-up construction for self-orthogonal codes,  introduce cyclic codes  and linear complementary dual (LCD) codes. 
We also gave a classification of self-orthogonal codes for short lengths.
\end{abstract}

\hspace{0.5in}

\section{Introduction}
\label{Sec:Intro}

Coding theory was studied classically using finite fields as alphabets. 
In the last thirty years, rings have been used along with finite fields, i.e., linear codes are defined as modules over a ring. 
Most research on codes over rings take commutative rings as alphabet \cite{SAS}, and in almost all cases, rings with unity (see \cite{Z4}, \cite{Zpm}, \cite{BU}, for example). Recently, there has been interest in non-unitary rings. One ring in particular is the non-local ring $H$ of order $4$ (see \cite{Fine}, \cite{Altassan.A et al.},\cite{Melaibari.A et al.})
The goal of this paper is to lay down the foundations of the study of duality of linear codes over two non-unital rings of order 6, denoted by $H_{23}$ and $H_{32}$ (see \cite{AAABBSS}), whose properties are similar to the ring $H$. These rings are interesting to study because a lot of new codes can be constructed over these rings. They also are in close connection to widely studied binary and ternary codes. For every code $\C$ over $H_{23}$ or $H_{32}$, there is an associated binary code $\C_a$ and a ternary code $\C_b$. This allows us to determine the characteristics of the code $\C$ by looking at the associated codes $\C_a$ and $\C_b$. In particular, conditions for self-orthogonal, self-dual, quasi self-dual and LCD codes are studied. Also, a method to construct self-orthgonal codes of longer length from codes of shorter length is presented as well as introduce cyclic codes over these rings.  We will also present a
classification of self-orthogonal codes for short lengths.

This paper is organized as follows. In Section \ref{sec:rings}, we define the rings $H_{23}$ and $H_{32}$ and give their properties. 
In Section \ref{sec:fields}, we recall some basic concepts on linear codes over finite field. In Section \ref{sec:duality}, we define codes over $H_{23}$ and $H_{32}$. We give characterizations of self-orthogonal, self-dual, quasi self-dual and LCD codes in terms of their associated binary and ternary codes. We also introduce cyclic codes over these rings. This is followed by a build-up construction, i.e. a method to construct codes from codes of smaller length in Section \ref{Sec:buildup}. In Section \ref{Sec:Results}, we  give a classification of self-orthogonal codes of lengths up to 7.

\section{The rings $H_{23}$ and $H_{32}$}\label{sec:rings}

We consider the rings $H_{23}$ and $H_{32}$, both of order $6$, given by the following presentation.
{\small \[ H_{23} = \langle a,b \mid 2a=0, 3b = 0, a^2=a, b^2 = 0, ab = 0 = ba \rangle,\]}
and
{\small \[H_{32} = \langle a,b \mid 2a=0, 3b = 0, a^2=0, b^2 = b, ab = 0 = ba \rangle. \]}
The other elements are defined as
\begin{eqnarray*}
c & := & a+b \\
d & := & 2b \\
e & := & a+2b
\end{eqnarray*}
The following table is the addition table for both rings,
and the multiplication tables for rings $H_{23}$ and $H_{32}$, respectively.
\begin{table}[ht]\label{tab:add}
\centering
\begin{tabular}{c|cccccc}
$+$ & $0$ & $a$ & $b$ & $c$ & $d$ & $e$ \\ \hline
$0$ & $0$ & $a$ & $b$ & $c$ & $d$ & $e$ \\
$a$ & $a$ & $0$ & $c$ & $b$ & $e$ & $d$ \\ 
$b$ & $b$ & $c$ & $d$ & $e$ & $0$ & $a$ \\
$c$ & $c$ & $b$ & $e$ & $d$ & $a$ & $0$ \\
$d$ & $d$ & $e$ & $0$ & $a$ & $b$ & $c$ \\
$e$ & $e$ & $d$ & $a$ & $0$ & $c$ & $b$ \\
\end{tabular}
\caption{Addition table for $H_{23}$ and $H_{32}$}
\end{table}

\begin{table}[ht]\label{tab:mult}
\centering
\begin{tabular}{c|cccccc}
$\cdot$ & $0$ & $a$ & $b$ & $c$ & $d$ & $e$ \\ \hline
$0$ & $0$ & $0$ & $0$ & $0$ & $0$ & $0$ \\
$a$ & $0$ & $a$ & $0$ & $a$ & $0$ & $a$ \\ 
$b$ & $0$ & $0$ & $0$ & $0$ & $0$ & $0$ \\
$c$ & $0$ & $a$ & $0$ & $a$ & $0$ & $a$ \\
$d$ & $0$ & $0$ & $0$ & $0$ & $0$ & $0$ \\
$e$ & $0$ & $a$ & $0$ & $a$ & $0$ & $a$ \\
\end{tabular}
\hspace{1cm}
\begin{tabular}{c|cccccc}
$\cdot$ & $0$ & $a$ & $b$ & $c$ & $d$ & $e$ \\ \hline
$0$ & $0$ & $0$ & $0$ & $0$ & $0$ & $0$ \\
$a$ & $0$ & $0$ & $0$ & $0$ & $0$ & $0$ \\ 
$b$ & $0$ & $0$ & $b$ & $b$ & $d$ & $d$ \\
$c$ & $0$ & $0$ & $b$ & $b$ & $d$ & $d$ \\
$d$ & $0$ & $0$ & $d$ & $d$ & $b$ & $b$ \\
$e$ & $0$ & $0$ & $d$ & $d$ & $b$ & $b$ \\
\end{tabular}
\caption{Multiplication table for $H_{23}$ and $H_{32}$, respectively}
\end{table}
\noindent From these tables, we see that both rings are commutative and without unity. They have two maximal ideals $J_a = \{0,a \}$ and $J_b = \{ 0,b,d \}$ and hence,  $H_{23}$ and $H_{32}$ are semi-local rings. From the definition, it is easy to verify that either of the two rings can be written as the direct sum $J_a \oplus J_b$. 

We define a natural action of $\mathbb{F}_2$ on
$J_a$ by the rule $r0 = 0r = 0$ and $r1 = 1r = r$, for all $r \in J_a$. This action is distributive in the sense that $r(s \oplus t) = r s+ rt$, where $\oplus$ denotes the addition in $\mathbb{F}_2$, $s,t \in \mathbb{F}_2$. We also define an action of $\mathbb{F}_3$ on $J_b$ in a similar fashion.
\section{Codes over finite fields}\label{sec:fields}

An $[n, k]$ linear code or an $[n, k]$-code $\C$ of length $n$ and dimension $k$ is a subspace of $\F^n$ where $p$ is a prime. The elements of such a code are called codewords. If $p=2$, then $\C$ is called binary code. If $p=3$, then $\C$ is called ternary code. The Hamming weight $wt(\x)$ of $\x \in \C$ is the number of nonzero coordinates in $\x$. We equip $\F^n$ with the standard inner product $(\x, \y)=\sum^n_{i=1}x_iy_i$ for $\x=(x_1,\ldots ,x_n)$,
$\y=(y_1,\ldots ,y_n)\in\F^n.$
The dual $\C^\perp$ of $\C$ is an $[n, n-k]$-code defined as
\begin{equation}
\C^\perp=\{\y \in\F^n \text{ } |\text{ }  (\x, \y)=0 \text{ } \text{for all} \text{ }  \x \in \C \}.
\end{equation} 
A linear code $\C$ is self-orthogonal if $\C\subseteq\C^\perp$ and self-dual if $\C=\C^\perp$. A linear code $\C$ is linear with complementary dual (LCD) if $\C \cap \C^\perp=\{\0\}$. A linear code $\C$  is cyclic provided that for each codeword $\c=(c_1,\cdots, c_n)$ in $\C$, the vector $(c_n,c_1,\cdots,c_{n-1} )$ is also in $\C$. 

\section{Duality of Codes over $H_{23}$ and $H_{32}$}\label{sec:duality}
Let $z\in\{23,32\}$. A linear  code $\C$ over $H_z$ or an $H_z$-code of length $n$ is an $H_z$-submodule of $H_z^n$.  
Since $H_z=J_a\oplus J_b, $ a code $\C$ over $H_{23}$ or $H_{32}$
can be written as the direct sum $\C=a\,\C_a\oplus b\,\C_b$,
where $\C_a$ is a binary code and $\C_b$ is a ternary code. 
 
The Hamming weight of a codeword $\c$ in an $H_z$- code $\C$, denoted $wt(\c)$, is the number of nonzero coordinates of $\c$ and the Hamming distance between codewords $\x$ and $\y$ in an $H_z$-code, denoted $d(\x,\y)$, is the number of coordinates where $\x$ and $\y$ are differ.   

Define the standard inner product on $H_z$ as 
\[(\x,\y)=\sum^n_{i=1} x_iy_i\] for all 
$\x=(x_1,\cdots, x_n), \y=(y_1, \cdots, y_n)\in H_z^n$. The dual $\C^{\perp} $of $\C$ is the submodule of $H_z^n$
defined by 
\begin{equation}
\C^\perp=\{\y \in\F^n \text{ } |\text{ }  (\x, \y)= 0 \text{ } \text{for all} \text{ }  \x \in \C \}.
\end{equation} 
An $H_z$-code $\C$ is self-orthogonal if $\C\subseteq\C^\perp$   and self-dual if $\C=\C^\perp$. An $H_z$- code $\C$ is said to be nice if $|\C||\C^\perp|=6^n$. If an $H_z$-code $\C$ is a self-orthogonal code and $|\C|=6^{\frac{n}{2}}$, we say $\C$ is a quasi self-dual (QSD) code.

\bigskip
\begin{thm}\label{thm:mindist}
Let $z \in \{23, 32 \}$. If $\C = a \, \C_a \oplus b \,\C_b$ is an $H_{z}$-code where $\C_a$ is a nonzero binary code of minimum distance $d_1$ and $\C_b$ is a nonzero ternary code of minimum distance $d_2$, then the  minimum distance of $\C$ is $d=\min\{d_1,d_2\}$. 
\end{thm}
\pf
Let $\u_1$ be a nonzero codeword of $\C_a$ of weight $d_1$ and $\u_2$ be a nonzero codeword of $\C_b$ of weight $d_2$. Since $a \,\C_a \subseteq \C$ and $b \,\C_b \subseteq \C$, this implies that,  $a\u_1, b\u_2 \in \C$ with $wt(a\u_1)=wt(\u_1)$ and  $wt(b\u_2)=wt(\u_2)$. Therefore, $d \leq \min\{d_1,d_2\}$. Conversely,  let $\u$ be a nonzero codeword of $\C$ of weight $d$. Then, $\u=a\x+b\y$ where $\x \in \C_a$ and $\y \in \C_b$. Since $\u\neq 0$, then one case of the following is true:
\begin{itemize}
\item If $\x =\0$ and $\y\neq 0$, then 
$d=wt(\u)=wt(b\y)=wt(\y)  \geq  d_2.$
\item If $\x \neq\0$ and $\y= 0$, then 
$d=wt(\u)=wt(a\x)=wt(\x)  \geq  d_1.$
\item If $\x \neq\0$ and $\y\neq 0$, then 
$d=wt(\u)\geq wt(a\y)=wt(\y)  \geq  d_2$ and 
$d=wt(\u)\geq wt(a\x)=wt(\x)  \geq  d_1.$
\end{itemize}
For all three cases, we have \[d=wt(\u)\geq \min\{d_1,d_2\},\]
and the result follows.
\qed
\begin{thm}\label{thmdualH23} 
Let $\C=a\, \C_a \oplus b\, \C_b$ be an $H_{z}$-code of length $n$. Then 
\begin{itemize}
\item[(i)] for $z=23$, $\C^{\perp} = a \, \C_a^\perp \oplus b \, \mathbb{F}_3^n$. Furthermore, $(\C^\perp)^\perp=\C$ if and only if $\C_b=\mathbb{F}_3^n$.
\item[(ii)] for $z=32$, $\C^{\perp} = a \, \mathbb{F}_3^n \oplus b \, \C_b^\perp $. Furthermore, $(\C^\perp)^\perp=\C$ if and only if $\C_a=\mathbb{F}_2^n$.
\end{itemize}
\end{thm}
\pf
Let $\C$ be an $H_{23}$-code and suppose $\u \in \C^{\perp}$. Write $\u=a\x+b\y$ where $\x \in \mathbb{F}^n_2$ and  $\y \in \mathbb{F}^n_3$. If  $\v\in \C_a$, then $a \v \in \C$.  Thus, we have
\begin{equation*}
0=(\u,a \v)=(a\x+b\y ,a \, \v)=a(\v,\x).
\end{equation*}  
Since $a \neq 0$, $(\v,\x)=0$,  so it follows that $\x \in \C_a^{\perp}$. Thus, we have $\u \in a \, \C_a^\perp + b \, \mathbb{F}_3^n$. Consequently, we have $\C^{\perp} \subseteq a \, \C_a^\perp + b \, \mathbb{F}_3^n$.

Now, suppose $\c \in a \, \C_a^\perp + b \, \mathbb{F}_3^n$, i.e., $\c = a\x +b\y$, where $\x \in  \C_a^\perp$ and $\y \in  \mathbb{F}_3^n$.  
For an arbitrary $\u=a\r +b\t\in\C$,  where $\r\in\C_a, \t\in\C_b$, we have 
\begin{equation*}
(\c,\u)=(a\x +b\y, a\r +b\t)=a(\x,\r)=a\cdot 0 =0.
\end{equation*} 
Since $\u$ is arbitrary, we see that $\c \in \C^{\perp}$ and hence, $a\,\mathcal{C}_a^\perp + b\, \mathbb{F}_3^n \subseteq \mathcal{C}^\perp$.
Then, we have $\C^{\perp} = a \, \C_a^\perp + b \, \mathbb{F}_3^n$. Since the intersection of the two vector spaces $a \, \C_a$ and $b \, \mathbb{F}_3^n$ is trivial, the sum is direct. Finally, since $(\C^\perp)^\perp=(a \, \C_a^\perp \oplus b \, \mathbb{F}_3^n)^\perp=a \, \C_a \oplus b \, \mathbb{F}_3^n$, we have $(\C^\perp)^\perp=\C$ if and only if $\C_b=\mathbb{F}_3^n$, so (i) follows. The case where $\C$ is an $H_{32}$ code is proved similarly.
\qed
\begin{cor}\label{nice} Let $\C=a\, \C_a \oplus b\, \C_b$ be an $H_{z}$-code of length $n$. 
Then $\C$ is nice if and only if 
\begin{enumerate}
\item[(i)] $\C_b=\{\mathbf{0}\}$ for $z=23$.
\item[(ii)] $\C_a = \{ \mathbf{0} \}$ for $z=32$.
\end{enumerate}

\end{cor}
\pf
Let $\C$ be an $H_{23}$ code and $k_a=\dim(\C_a)$ and $k_b=\dim(\C_b)$. Then $|\C|=2^{k_a}3^{k_b}$. 
By Theorem~\ref{thmdualH23} (i),  $|\C^{\perp}|=2^{n-k_a}3^{n}$.
Thus, 
\[|\C||\C^{\perp}|=2^{n}\,3^{n+k_b}.\] 
Therefore, $\C$ is nice if and only if $k_b=0$. 

Similarly, (ii) follows from Theorem~\ref{thmdualH23} (ii).  \qed
\subsection{Self-orthogonal, Self-dual and QSD codes}
We now determine the conditions so that a code $\C$ over $H_{23}$ or $H_{32}$ is self-orthognal and quasi self-dual.

\begin{thm}\label{thmorthH23} 
Let $\C=a\, \C_a \oplus b\, \C_b$ be an $H_{z}$-code. 
\begin{enumerate}
\item[(i)] If $z=23$, then $\C$ is self-orthogonal if and only if $\C_a$ is a binary self-orthogonal code.
\item[(ii)] If $z=32$, then $\C$ is self-orthogonal if and only if $\C_b$ is a ternary self-orthogonal code.
\end{enumerate}
\end{thm}
\pf
Let $\C$ be an $H_{23}$ code and suppose $\c_1=a\x_1+by_1$ and $\c_2=a\x_2+b\y_2$ are arbitrary codewords in $\C$ where $\x_1$, $\x_2\in \C_a$ and $\y_1$, $\y_2\in\C_b$. 
We have
\begin{equation}
\label{eqn:1}
\begin{split}
(\c_1,\c_2) = & a^2(\x_1,\x_2) + b^2 (\y_1,\y_2)\\
= & a(\x_1,\x_2).
\end{split}
\end{equation}
 Since $a \neq 0$,  by (\ref{eqn:1}), $\C$ is a self-orthogonal if and only if $\C_a$ is a binary self-orthogonal code.
 
 On the other hand, $a^2 = 0$ and $b^2 = b$ in $H_{32}$, so (ii) follows.
%
\qed


\begin{thm}\label{thmbinaryternary} Let $\C_1$ be a binary linear code and $\C_2$ be a ternary linear code.
The code $\C$ defined by
\[\C=a\,\C_1\oplus b\,\C_2\] 
is a self-orthogonal $H_{z}$-code of length $n$ where $\C_a=\C_1$ and $\C_b=\C_2$ if the following conditions are satisfied:
\begin{enumerate}
\item[(i)] For $z=23$, $\C_1$ is a binary self-orthogonal code of length $n$ and $\C_2$ is a ternary code of length $n$.
\item[(ii)] For $z=32$, $\C_1$ is a binary code of length $n$ and $\C_2$ is a ternary self-orthogonal code of length $n$.
\end{enumerate}
Moreover, if $|\C_1||\C_2|=6^{n/2}$, then $\C$ is QSD.
\end{thm}
\pf Let $z=23$. By the linearity of $\C_1$ and $\C_2$,  the code $\C=a\, \C_1 \oplus b\, \C_2$ is closed under addition.  
For the scalar multiplication,  let $\c=a\x+b\y\in\C$ where $\x\in\C_1$ and $\y\in\C_2$. Then we have the following:
\begin{itemize}
\item $a(a\x+b\y) = a^2\x = a\x =a\x + b\mathbf{0} \in\C$;
\item $b(a\x+b\y) = \mathbf{0} = a\mathbf{0} +b\mathbf{0} \in\C$.
\end{itemize}
Since $\C$ is closed under addition,  for any $r=ax+by \in H_{23}$ where $x\in \mathbb{F}_2$ and $y\in \mathbb{F}_3$,
we have $r\,\C \subseteq \C$. Hence, $\C$ is a linear code over $H_{23}$.
To prove the self-orthogonality of $\C$, for all $a\x_1+b\y_1,a\x_2+b\y_2\in\C$ where $\x_1, \x_2\in \C_1$  and $\y_1,\y_2 \in \C_2$, we have 
\begin{equation*}
\begin{split}
(a\x_1+b\y_1,a\x_2+b\y_2) & = a^2(\x_1,\x_2) + b^2 (\y_1,\y_2) \\
& =a(\x_1,\x_2) \\
& = 0,
\end{split}
\end{equation*}
since $\C_1$ is a binary self-orthogonal code. 

Statement (ii) is proved similarly.

 Furthermore,  since $|\C|=|\C_1||\C_2|$,  the last statement holds. \qed


\begin{cor}\label{H23QSD}
Let $\C=a\, \C_a \oplus b\, \C_b$ be an $H_{z}$-code of length $n$. Then  $\C$ is a QSD code of length $n$ if and only if the following conditions are satisfied:
\begin{enumerate}
\item[(i)] For $z=23$, $\C_a$ is a binary self-dual code and $\C_b$ is a ternary $[n,n/2]$ code.
\item[(ii)] For $z=32$, $\C_a$ is a binary $[n,n/2]$ code and $\C_b$ is a ternary self-dual code.
\end{enumerate}
\end{cor}
\pf
The result follows from \cite[Lemma 1]{AAABBSS},  Theorem~\ref{thmorthH23}, Theorem~\ref{thmbinaryternary}
and the definition of QSD codes. 
\qed

\begin{thm}\label{SD}
Let $\C=a\, \C_a \oplus b\, \C_b$ be an $H_{z}$-code of length $n$. Then  $\C$ is self-dual if and only if the following conditions are satisfied:
\begin{enumerate}
\item[(i)] For $z=23$, $\C_a$ is a binary self-dual  code and $\C_b=\mathbb{F}_3^n$.
\item[(ii)] For $z=32$, $\C_a = \mathbb{F}_2^n$ and $\C_b$ is a ternary self-dual code.
\end{enumerate}
\end{thm}
\pf
Let $\C$ be an $H_{23}$-code. From the definition of self-dual codes and Theorem \ref{thmdualH23} (i), we have
\begin{equation*}
 a\, \C_a \oplus b\, \C_b=\C=\C^\perp=a \, \C_a^\perp \oplus b \, \mathbb{F}_3^n.
\end{equation*}
 Therefore, $\C$ is self-dual if and only if $\C_a=\C_a^\perp$ and $\C_b=\mathbb{F}_3^n$. Similarly, (ii) follows from the definition of self-dual codes and Theorem \ref{thmdualH23} (ii).
\qed

\bigskip
Observe that by Corollary~\ref{H23QSD} and Theorem~\ref{SD},  QSD and self-dual $H_{23}$-codes have even lengths while QSD and self-dual $H_{32}$-codes have length divisible by 4 since $\C_b$ must be ternary self-dual codes.

Now, we will show that any $H_{23}$-code cannot be both QSD and self-dual at the same time. 
\begin{prop}\label{QSDnotSD}
Let $\mathcal{Q}$ be the set of all QSD codes of length $n$ over $H_{23}$ and $\mathcal{S}$ be the set of all self-dual
codes of length $n$ over $H_{23}$. Then $\mathcal{Q}\cap \mathcal{S}=\emptyset$.
\end{prop}
\pf
Suppose $\C$ is an $H_{23}$-code of length $n$ that is both QSD and self-dual. 
		Then $|\C|= |\C^{\perp}|= 6^{n/2}$. Therefore, $|\C||\C^{\perp}|= 6^{n}$, 
		so $\C$ is a nice code.  However, by Corollary \ref{nice},  $\C_b=\{\mathbf{0}\}$
		which contradicts Corollary~\ref{H23QSD}.
		Thus, $\mathcal{Q}\cap \mathcal{S}= \emptyset$.
\qed
The same can be said for $H_{32}$ codes.
\subsection{Cyclic codes}
A cyclic code $\C$ of length $n$ over $H_z$ is a linear code such that if $(c_1, \cdots , c_n) \in \C$, then its cyclic shift $(c_n,c_1,\cdots, c_{n-1})$ is also a codeword in $\C$.

\begin{thm}\label{thmcyclic}
Let $\C=a\, \C_a \oplus b\, \C_b$ be an $H_{z}$-code of length $n$, where $z \in \{23, 32 \}$. Then $\C$ is a cyclic code if and only if $\C_a$ and  $\C_b$ are both cyclic.
\end{thm}
\pf
Suppose that $\C$ is a cyclic code over $H_{z}$. Let $(x_1,\ldots,x_n)\in \C_a$ and  $(y_1,\ldots,y_n)\in \C_b$. Then $(ax_1+by_1,\ldots ,ax_n+by_n)$ is a codeword in $\C$.  Since $\C$ is closed under cyclic shifts,  
\begin{equation*}
\begin{split}
& a( x_n, x_1,\ldots ,x_{n-1}) +b(y_n,y_1,\ldots ,y_{n-1}) \\
& =(a x_n+b y_n, ax_1+by_1,\ldots ,ax_{n-1}+by_{n-1})\in\C.
\end{split}
\end{equation*}
Therefore $( x_n, x_1,\ldots ,x_{n-1})$ and $(y_n,y_1,\ldots ,y_{n-1})$ are codewords from $\C_a$ and $\C_b$, respectively. 
Thus,  $\C_a$ and $\C_b$ are both closed under cyclic shifts, hence they are cyclic codes. Now, we prove the reverse direction. Since $\C$ is defined as $\C = a\,C_a \oplus b\,C_b$, any linear combination of codewords from $\C_a$ and $\C_b$ will be a codeword in $\C$.  Since both $\C_a$ and $\C_b$ are cyclic codes, it follows that $\C$ is a cyclic code.
 \qed
  \begin{cor}
  If $\C=a\, \C_a \oplus b\, \C_b$ is a cyclic code over $H_{z}$, where $z \in \{23, 32 \}$, then its dual is also cyclic. 
  \end{cor}
  \pf
  The result follows from Theorem \ref{thmdualH23} and Theorem \ref{thmcyclic}.   
  \qed
 \begin{cor}\label{cyclic}
 A QSD code $\C=a\, \C_a \oplus b\, \C_b$ of even length $n$ over $H_{z}$ is cyclic if and only if the following conditions are satisfied:
 \begin{enumerate}
 \item[(i)] For $z = 23$, $\C_a$ is a binary cyclic self-dual code and $\C_b$ is a ternary cyclic  $[n,n/2]-$code.
 \item[(ii)] For $z=32$, $\C_a$ is a binary cyclic  $[n,n/2]-$code and $\C_b$ is a ternary cyclic self-dual code.
 \end{enumerate}
  \end{cor}
  \pf The result follows from  Corollary \ref{H23QSD} and Theorem \ref{thmcyclic}.   
  \qed
 
\subsection{LCD codes}
An $H_z$-code $\C$ is linear with complementary dual (LCD) if $\C \cap \C^\perp=\{\0\}$. The following theorem gives a characterization of LCD codes over the ring $H_z$.

\bigskip
\begin{thm}\label{LCD}
A linear code $\C = a\,\C_a + b\,\C_b$ over $H_{z}$ is LCD if and only if the following conditions are satisfied:
\begin{itemize}
\item[(i)] For $z=23$, $\C_a$ is LCD and $\C_b = \{ \mathbf{0} \}$ iff $\C$ is nice.
\item[(ii)] For $z=32$, $\C_a = \{ \mathbf{0} \}$ and $\C_b$ is LCD iff $\C$ is nice.
\end{itemize}
\end{thm}

\pf 
Let $z=23$ and suppose $\C = a\,\C_a + b\,\C_b$ is LCD, i.e., $\C \cap \C^\perp = \{ \mathbf{0} \}$. By Theorem \ref{thmdualH23}, 
$ \C_a \cap \C_a^\perp = \{ \mathbf{0}  \}$ and $\C_b  \cap \mathbb{F}_3^n = \{ \mathbf{0}  \}$. This implies $\C_a$ is LCD and $\C_b = \{ \mathbf{0}  \}$. By Corollary \ref{nice}, $\C$ is nice.

Suppose $\C_a$ is LCD and $\C_b = \{\mathbf{0} \}$. By definition, $\C_a \cap \C_a^\perp = \{\mathbf{0} \}$. 
So we have $\C \cap \C^\perp = a(\C_a \cap \C_a^\perp) + b (\C_b \cap \C_b^\perp) = \{\mathbf{0} \}$, which shows $\C$ is LCD.

The case where $\C$ is an $H_{32}$-code is proved similarly.
\qed

\begin{cor}\label{cor:LCD}
Suppose $\C= a\,\C_a + b\,\C_b$  is a  cyclic code  over $H_{z}$. 
Then $\C$ is an LCD if and only if the following conditions are satisfied:
\begin{itemize}
\item[(i)] For $z=23$, $\C_a$ is an LCD cyclic code and $\C_b = \{ \mathbf{0} \}$.
\item[(ii)] For $z=32$, $\C_a = \{ \mathbf{0} \}$ and $\C_b$ is an LCD cyclic code.
\end{itemize}
\end{cor}

\pf 
The result follows from Theorem \ref{thmcyclic} and  Theorem \ref{LCD}.   
\qed
\section{Build-up construction for self-orthogonal codes}\label{Sec:buildup}

In this chapter, we give several methods to construct self-orthogonal codes from codes of smaller length. 
These are called the {\it building up constructions}. 

Recall the action of $\mathbb{F}_2$ on $J_a$ and $\mathbb{F}_3$ on $J_b$. If $\x = (x_1, \ldots, x_n) \in \mathbb{F}_2^n$, define $\alpha\x = (\alpha x_1, \ldots, \alpha x_n) \in H_{23}^n$ for $\alpha \in J_a$. Similarly, if $\x = (x_1, \ldots, x_n) \in \mathbb{F}_3^n$, define $\beta \x = (\beta x_1, \ldots, \beta x_n) \in H_{32}^n$ for $\beta \in J_b$. 

\begin{thm}\label{thm:H23build1}
Let $\C_0$ be a self-orthogonal code over $H_{23}$ of length $n$ 
with generating set $\{\r_1,\ldots ,\r_k\}$.  Let $\x = (x_1, \ldots, x_n)$ be a fixed binary vector satisfying $(\x,\x)=1$, that is, $\x$ is a vector of odd Hamming weight. Denote $y_i=(\x,\r_i),$ $1\leq i\leq k$. Then the code $\C$ with $k+1$ generators given by
\begin{equation}
(\alpha,0,\alpha \x ), \, (y_1,y_1,\r_1),\, \ldots , \, (y_k,y_k,\r_k),
\end{equation}
where $\alpha\in J_a$ is a self-orthogonal code  over $H_{23}$ of length $n+2$.
\end{thm}
\pf 
\begin{itemize}
\item The first vector is orthogonal to itself since  $\alpha^2+(\x,\x) \cdot \alpha^2 =\alpha+\alpha=0$.
\item The first vector is orthogonal to the other $k$ vectors since $\alpha y_i + (\alpha \x,\r_i)=2\alpha y_i=0.$
\item The last $k$ vectors are orthogonal to each other and to themselves by self-orthogonality of $\C_0$ 
since $y_iy_j +y_iy_j +(\r_i,\r_j)=2y_iy_j + 0=0$,
where $y_iy_j\in J_a$ in $H_{23}$.  
\end{itemize}
Hence $\C$ is a self-orthogonal code over $H_{23}$ of length $n+2$.
\qed

\begin{thm}\label{thm:H23build2}
Let $\C_0$ be a self-orthogonal code over $H_{23}$ of length $n$ with generating set $\{\r_1,\ldots ,\r_k\}$. 
Let $\x = (x_1, \ldots, x_n)$ be a fixed  ternary vector.  Denote $y_i=(\x,\r_i),$ $1\leq i\leq k$. 
Then the code $\C$ with $k+3$
generators given by
\begin{equation}
\begin{split}
(\beta,0,0,\beta\x), (0,\beta,0,\beta\x), (0,0,\beta,\beta\x),\\
 (2y_1,2y_1,2y_1,\r_1),\ldots ,(2y_k,2y_k,2y_k,\r_k),
 \end{split}
\end{equation}
where $\beta\in J_b=\{0,b,d\}$ is a self-orthogonal code over $H_{23}$ of length $n+3$ .
\end{thm}
\pf
\begin{itemize}
\item The first three vectors are orthogonal to itself since  $\beta^2+\beta^2(\x,\x)=0+0(\x,\x)=0$. 
\item The first three vectors are orthogonal to each other as  $\beta^2(\x,\x)=0$. 
\item The first three vectors are orthogonal to any of the last $k$ vectors since 
$2\beta y_i+(\beta\x,\r_i)=3\beta y_i=0.$
\item The last $k$ rows are orthogonal to each other and to themselves by self-orthogonality of $\C_0$ 
since 
$4y_iy_j +4y_iy_j + 4y_iy_j+(\r_i,\r_j)=12y_iy_j+0=0$.  
\end{itemize}
Hence $\C$ is a self-orthogonal code over $H_{23}$ of length $n+3$.
\qed
\begin{ex}
Consider the self-orthogonal code $\C_0$ of length 2 over $H_{23}$ generated by vectors 
\begin{equation*}
\r_1= (a,a)  \,  \text{ and }  \, \r_2=(b,0). 
\end{equation*}

By using  Theorem \ref{thm:H23build1}, we will construct a self-orthogonal code over $H_{23}$ of length 4 and 6 from $\C_0$.
Let $\alpha=a$ and $\x=(1,0) \in \mathbb{F}_2^2$.  Then $y_1=a$ and $y_2=b$. we obtain a new self-orthogonal code over $H_{23}$
of length $4$ and with the following generators given by $(a,0,a,0 ), \,(a,a,a,a), \, (b,b,b,0)$.
Let 
\begin{equation*}
\r'_1=(a,0,a,0 ), \, \r'_2=(a,a,a,a), \, \r'_3=(b,b,b,0).
\end{equation*}
Now, by using $\r'_1, \r'_2$ and $\r'_3$ with $\x=(1,0,1,1)\in \mathbb{F}^2_2$ and $\alpha=a$,
we have $y_1=0$, $y_2=a$ and $y_3=d$. 
We obtain a self-orthogonal code over $H_{23}$ of length $6$ with generators given by
\begin{equation*}
(a,0,a,0,a,a), \, (0,0,a,0,a,0), \, (a,a,a,a,a,a),
    (d,d,b,b,b,0)
\end{equation*}

By using Theorem \ref{thm:H23build2} with $\beta=b$ and $\x=(2,1) \in \mathbb{F}^2_3$, 
we will construct a self-orthogonal code over $H_{23}$ of length 5 and 8 from $\C_0$.
 Then $y_1=a$ and $y_2=d$. We obtain a self-orthogonal code of length $5$ with generators given by
\begin{equation*}
\g_1=(b,0,0,d,b ), \, \g_2=(0,b,0,d,b), \, \g_3=(0,0,b,d,b),
 \, \g_4=(0,0,0,a,a), \, \g_5=(b,b,b,b,0).
\end{equation*}
Now,  by using $\g_1, \g_2, \g_3,\g_4$ and $\g_5$, with $\x=(1,0,0,2,2)\in \mathbb{F}^5_3$ and $\beta=d$,
we will construct a self-orthogonal code over $H_{23}$ of length 8. Then, we have $y_1=b$, $y_2=y_3=y_4=y_5=0$. 
We obtain a self-orthogonal code over $H_{23}$ of length $8$ with generators given by
\begin{equation*}
(d,0,0,d,0,0,b,b ), \, (0,d,0,d,0,0,b,b), \, (0,0,d,d,0,0,b,b),\, (d,d,d,b, 0,0,d,b) ,
 \end{equation*}
 \begin{equation*}
\, (0,0,0,0,b,0,d,b), \, (0,0,0,0,0,b,d,b), \,(0,0,0,0,0,0,a,a), \,(0,0,0,b,b,b,b,0).
\end{equation*}

\end{ex}
\begin{thm}\label{thm:H32build}
Let $\C_0$ be a self-orthogonal code of length $n$ over $H_{32}$  with generating set $\r_1,\ldots ,\r_k$.  
Let $\x$ be a fixed  ternary vector of length $n$ and $\alpha, \beta,$ and $
\gamma \in H_{32}$ such that $\alpha+ \beta + \gamma=0. $ Denote $y_i=(\x,\r_i),$ $1\leq i\leq k$. 
Then the code $\C$ with $k+1$
generators given by
\begin{equation}
(\alpha,\beta,0,\gamma\x), (y_1,y_1,y_1,\r_1),\ldots ,(y_k,y_k,y_k,\r_k),
\end{equation}
is a self-orthogonal code over $H_{32}$ of length $n+3$, if one of the following hold:
\begin{itemize}
\item[(i)] $(\x,\x)=1$ and $\alpha^2+ \beta^2 + \gamma^2=0$,
\item[(ii)] $(\x,\x)=-1$ and $\alpha^2+ \beta^2 - \gamma^2=0.$\end{itemize}
\end{thm}
\pf

\begin{enumerate}
\item[(i)] If $(\x,\x)=1$ and $\alpha^2+ \beta^2 + \gamma^2=0$, then
\begin{itemize}
\item The first vector is orthogonal to itself since  $\alpha^2+ \beta^2 + \gamma^2 (\x,\x)=\alpha^2+ \beta^2 + \gamma^2=0$.
\item The first vector is orthogonal to any of the other $k$ vectors since 
$\alpha y_i+\beta y_i+\gamma y_i=(\alpha+ \beta + \gamma)y_i=0.$
\item The last $k$ vectors are orthogonal to each other and to themselves by self-orthogonality of $\C_0$ since 
$3y_iy_j+(\r_i ,\r_j)=0$,
where $y_iy_j\in J_b$ in $H_{32}$.  
\end{itemize}
\item[(ii)] If $(\x,\x)=-1$ and $\alpha^2+ \beta^2 - \gamma^2=0,$ then
\begin{itemize}
\item The first vector is orthogonal to itself since   $\alpha^2+ \beta^2 + \gamma^2 (\x,\x)=\alpha^2+ \beta^2 - \gamma^2 =0$.
\item The first vector is orthogonal to any of the other $k$ vectors since $\alpha y_i+\beta y_i+\gamma y_i=(\alpha+ \beta + \gamma)y_i=0.$
\item The last $k$ vectors  are orthogonal to each other and to themselves by self-orthogonality of $\C_0$ since 
$3y_iy_j +(\r_i ,\r_j)=0$,
where $y_iy_j\in J_b$ in $H_{32}$.  
\end{itemize}
\end{enumerate}
Hence $\C$ is a self-orthogonal code over $H_{32}$ of length $n+3$.\qed
\begin{ex}
By using  Theorem \ref{thm:H32build}, we will construct a self-orthogonal code over $H_{32}$ 
of length 6 and 9  from a self-orthogonal code $\C_0$ of length 3 generated by

\begin{equation*}
\r_1= (a,0,0)\, \text{ and } \,\r_2=(b,b,b). 
\end{equation*}

Let $\alpha=a, \, \beta=b, \, \gamma=e,$ and  $\x=(0,1,1) \in \mathbb{F}_3^3$.  Then $y_1=0$ and $y_2=d$. 
By Theorem \ref{thm:H32build} (ii), we obtain a 
self-orthogonal code over $H_{32}$ of length $6$ with generators given by
\begin{equation*}
\r'_1=(a,b,0,0,e,e ), \, \r'_2=(0,0,0,a,0,0), \, \r'_3=(d,d,d,b,b,b).
\end{equation*}

Consider the self-orthogonal code
generated by $\r'_1, \r'_2$ and $\r'_3$.
Let $\alpha=e ,\, \beta=e ,\, \gamma=d$ and  $\x=(2,2,0,1,1,0) \in \mathbb{F}_3^6$.  Then $y_1=c, \, y_2=a$ and $y_3=b$. 
By Theorem \ref{thm:H32build} (i), we obtain a self-orthogonal code over $H_{32}$ of length $9$ with generators given by
\begin{equation*}
(e,e,0,b,b,0,d,d,0), \, (c,c,c,a,b,0,0,e,e), \, (a,a,a,0,0,0,a,0,0),\, (b,b,b,d,d,d,b,b,b).
\end{equation*}
\end{ex}
\section{Numerical results}
\label{Sec:Results}

In this section, we give a classification of self-orthogonal codes over $H_{23}$ and $H_{32}$ of lengths up to 7.
We a method similar to \cite{AAABBSS}. For each pair $(\C_a, \C_b)$, we find self-orthogonal codes that are permutation
equivalent to $\C = a\,\C_a+b\,\C_b$. This can be accomplished by double coset decompositions of the symmetry group $S_n$ by automorphism groups of codes $\C_a$ and codes $\C_b$ satisfying conditions given in Theorem \ref{thmbinaryternary}. Then we take the System of Distinct Representatives (SDR), i.e., a set of representative of the double cosets in the group $S_n$. Here, we have an analog of Theorem 2 in \cite{AAABBSS}.

\begin{thm}\label{H23SDR}
Let $(\C_a,\C_b)$ be a pair of codes as defined in Theorem \ref{thmbinaryternary}. 
Let $\Aut(\C_a)$ and $ \Aut(\C_b)$ denote permutation groups of $\C_a$ 
and $\C_b$, respectively. Then the set
\begin{equation*} 
S_{\C_a,\C_b} := \{ a \, \C_a + b \, \sigma (\C_b) \, | \,  \sigma \text{ runs over an SDR of }
 \Aut(\C_a) \backslash S_n / \Aut(\C_b) \} 
\end{equation*}
forms a set of permutation inequivalent self-orthogonal $H_{z}$-codes. Moreover, $|S_{\C_a,\C_b} |= |Aut(\C_a) \backslash S_n /Aut(\C_b)|$.
\end{thm}

To carry out the classification, we note of the following consequence of Theorem \ref{H23SDR}, similar to Corollary 1 and 
Corollary 2 in \cite{AAABBSS}.

\begin{cor}
Let $L_a$ be the set of all inequivalent binary codes of length $n$ and $L_b$ be
the set of all (permutation) inequivalent ternary codes that satisfies Theorem  \ref{thmbinaryternary}.  
Then the set of all self-orthogonal $H_{z}$-codes of length
$n$ is the disjoint union
\[ \bigcup_{\C_a \in La, \, \C_b \in L_b} S_{\C_a,\C_b}.\]
\end{cor}

Based on the above results, we have the classification algorithm for self-orthogonal $H_{z}$-codes of length $n$ as follows:
\begin{enumerate} 
\item If $z=23$, 
\begin{enumerate}
\item write a list $L_a$ of inequivalent binary self-orthogonal $[n,k_a]$-codes;
\item write a list $L_b$ of permutation inequivalent ternary $[n,k_b]$-codes.
\end{enumerate}
\item If $z=32$, 
\begin{enumerate}
\item write a list $L_a$ of inequivalent binary $[n,k_a]$-codes;
\item write a list $L_b$ of permutation inequivalent ternary self-orthogonal $[n,k_b]$-codes.
\end{enumerate}
\item For every pair $(\C_a,\C_b) \in L_a \times L_b$,
	\begin{enumerate}
	\item compute the automorphism groups $Aut(\C_a)$ and $Aut(\C_b)$,
	\item determine a list $\sigma_1, \ldots, \sigma_r$ of representatives of $Aut(\C_a) \backslash S_n / Aut(\C_b)$,
	\item for $i= 1, \ldots, r$, compute $a \, \C_a + b \, \sigma_i (\C_b)$.
	\end{enumerate}
\end{enumerate}
Clearly, the zero code of length $n$ is self-orthogonal.
The following results are for nonzero codes. All computations are done with the help of 
MAGMA \cite{MAGMA}.  
\subsection{Self-orthogonal codes over $H_{23}$}
If $(k_a,k_b)=(0,n)$, then we have the self-orthogonal code $\C=b\mathbb{F}_3^n$.
If $(k_a,k_b)=(0,k_b)$, then the number of inequivalent self-orthogonal codes over $H_{23}$
for a given length $n$ is equal to the number of permutation inequivalent ternary codes of dimension $k_b$.
If $(k_a,k_b)=(k_a,0)$ or $(k_a,k_b)=(k_a,n)$, then the number of inequivalent self-orthogonal codes over $H_{23}$
for a given length $n$ is equal to the number of inequivalent binary self-orthogonal codes of dimension $k_a$.
For the other cases,  please refer to Table 3.
\begin{table}[ht]\label{tab:classifyH23}
\caption{Number of nonzero inequivalent self-orthogonal $H_{23}$-codes of lengths $n$ up to $7$}
\begin{center}

{\footnotesize 
\begin{tabular}{cccccc|cccccc}
\hline 
$n$ & $(k_{a},k_{b})$ & $|L_a|$ &  $|L_b|$ & $\#$Codes & Remark & $n$ & $(k_{a},k_{b})$ & $|L_a|$ &  $|L_b|$ & $\#$Codes & Remark \tabularnewline
\hline 
$2$ & $(1,1)$ & $1$ &  $3$ & 3 & QSD  & $6$ & $(2,4)$ & $3$ &  $87$ & $2003$ &  \tabularnewline
\cline{1-6} \cline{2-6} \cline{3-6} \cline{4-6}  \cline{5-6} \cline{6-6} 
$3$ & $(1,1)$ & $1$ &  $5$ & 9 & & $6$ & $(2,5)$ & $3$ &  $15$ & $145$ &  \tabularnewline
$3$ & $(1,2)$ & $1$ &  $5$ & 9 & & $6$ & $(3,1)$ & $1$ &  $15$ & $31$ &  \tabularnewline
\cline{1-6} \cline{2-6} \cline{3-6} \cline{4-6}  \cline{5-6} \cline{6-6} 
$4$ & $(1,1)$ & $2$ &  $8$ & 27 & &  $6$ & $(3,2)$ & $1$ &  $87$ & $404$ &  \tabularnewline
$4$ & $(1,2)$ & $2$ &  $16$ & 66 & & $6$ & $(3,3)$ & $1$ &  $168$ & $1032$ & QSD  \tabularnewline
$4$ & $(1,3)$ & $2$ &  $8$ & 27 & & $6$ & $(3,4)$ & $1$ &  $87$ & $404$ &  \tabularnewline
$4$ & $(2,1)$ & $1$ &  $8$ & 12 & & $6$ & $(3,5)$ & $1$ &  $15$ & $31$ &  \tabularnewline
\cline{7-12} \cline{8-12} \cline{9-12} \cline{10-12}  \cline{11-12} \cline{12-12} 
$4$ & $(2,2)$ & $1$ &  $16$ & 30 & QSD & $7$ & $(1,1)$ & $3$ &  $19$ & $185$ &  \tabularnewline
$4$ & $(2,3)$ & $1$ &  $8$ & 12 &  & $7$ & $(1,2)$ & $3$ &  $176$ & $3822$ &  \tabularnewline 
\cline{1-6} \cline{2-6} \cline{3-6} \cline{4-6}  \cline{5-6} \cline{6-6} 
$5$ & $(1,1)$ & $2$ &  $11$ & 54 &  & $7$ & $(1,3)$ & $3$ &  $644$ & $20458$ &  \tabularnewline
$5$ & $(1,2)$ & $2$ &  $39$ & $289$ &  & $7$ & $(1,4)$ & $3$ &  $644$ & $20458$ &  \tabularnewline
$5$ & $(1,3)$ & $2$ &  $39$ & $289$  &  & $7$ & $(1,5)$ & $3$ &  $176$ & $3822$ &  \tabularnewline
$5$ & $(1,4)$ & $2$ &  $11$ & $54$  &  & $7$ & $(1,6)$ & $3$ &  $19$ & $185$ & \tabularnewline
$5$ & $(2,1)$ & $1$ &  $11$ & $33$  &  & $7$ & $(2,1)$ & $3$ &  $19$ & $333$ &  \tabularnewline
$5$ & $(2,2)$ & $1$ &  $39$ & $220$  &  & $7$ & $(2,2)$ & $3$ &  $176$ & $11400$ &  \tabularnewline

$5$ & $(2,3)$ & $1$ &  $39$ & $220$  &  & $7$ & $(2,3)$ & $3$ &  $644$ & $76968$ &  \tabularnewline
$5$ & $(2,4)$ & $1$ &  $11$ & $33$  &  & $7$ & $(2,4)$ & $3$ &  $644$ & $76968$ &  \tabularnewline
$6$ & $(1,1)$ & $3$ &  $15$ & $109$  &  & $7$ & $(2,5)$ & $3$ &  $176$ & $11400$ &  \tabularnewline
$6$ & $(1,2)$ & $3$ &  $87$ & $1143$ &  & $7$ & $(2,6)$ & $3$ &  $19$ & $333$ &  \tabularnewline
$6$ & $(1,3)$ & $3$ &  $168$ & $2640$ &  & $7$ & $(3,1)$ & $2$ &  $19$ & $118$ &  \tabularnewline
$6$ & $(1,4)$ & $3$ &  $87$ & $1143$ &  & $7$ & $(3,2)$ & $2$ &  $176$ & $4074$ &  \tabularnewline
$6$ & $(1,5)$ & $3$ &  $15$ & $109$ &  & $7$ & $(3,3)$ & $2$ &  $644$ & $29998$ &  \tabularnewline
$6$ & $(2,1)$ & $3$ &  $15$ & $145$  &  & $7$ & $(3,4)$ & $2$ &  $644$ & $29998$ &  \tabularnewline
$6$ & $(2,2)$ & $3$ &  $87$ & $2003$ &  & $7$ & $(3,5)$ & $2$ &  $176$ & $4074$ &  \tabularnewline
$6$ & $(2,3)$ & $3$ &  $168$ & $5096$ & & $7$ & $(3,6)$ & $2$ &  $19$ & $118$ &  \tabularnewline

\hline
\end{tabular}}
\end{center}
\end{table}

There is a unique binary self-dual code of length $2$ with generator matrix \[\left[ \begin{array}{cc} 1 & 1 \end{array} \right].\] Up to equivalence, we find $3$ QSD codes of length $2$ over $H_{23}$ and each of their associated codes $\C_a$ and $\sigma_i(\C_b)$ are listed below. By Corollary \ref{cyclic}, all these codes are cyclic. 
\begin{center}
{\small \begin{tabular}{|c|c|c|}
\hline
$\C_a$ & $\sigma_i(\C_b)$ & Remark \\ 
\hline \hline
$[1,1]$ & $[1,0]$ & cyclic\\
\hline
& $[1,1]$ & cyclic \\ 
\hline
& $[1,2]$ & cyclic \\
\hline
\end{tabular} }
\end{center} 

There is a unique binary self-dual code $\C_a$ of length $4$ with generator matrix 
\[G_a=\left[ \begin{array}{cccc} 1 & 0 & 1 & 0 \\ 0 & 1 & 0 & 1 \end{array} \right].\] Up to equivalence, we find $30$ inequivalent 
QSD codes of length $4$ over $H_{23}$ and associated ternary codes $\sigma_i(\C_b)$ has generator matrices listed below:
$$ \left[ \begin{array}{cccc} 1 & 2 & 2 & 0 \\ 0 & 0 & 0 & 1 \end{array} \right],  \left[ \begin{array}{cccc} 1 & 2 & 1 & 0 \\ 0 & 0 & 0 & 1 \end{array} \right], \left[ \begin{array}{cccc} 1 & 2 & 0 & 0 \\ 0 & 0 & 1 & 2 \end{array} \right], 
\left[ \begin{array}{cccc} 1 & 0 & 2 & 0 \\ 0 & 1 & 0 & 2 \end{array} \right], \left[ \begin{array}{cccc} 1 & 0 & 0 & 2 \\ 0 & 1 & 0 & 2 \end{array} \right], $$
$$ \left[ \begin{array}{cccc} 1 & 0 & 2 & 0 \\ 0 & 1 & 0 & 0 \end{array} \right],  \left[ \begin{array}{cccc} 1 & 0 & 0 & 0 \\ 0 & 1 & 2 & 0 \end{array} \right], \left[ \begin{array}{cccc} 1 & 0 & 2 & 2 \\ 0 & 1 & 0 & 1 \end{array} \right], 
\left[ \begin{array}{cccc} 1 & 0 & 0 & 1 \\ 0 & 1 & 2 & 2 \end{array} \right], \left[ \begin{array}{cccc} 
1 & 0 & 1 & 2 \\ 0 & 1 & 1 & 0 \end{array} \right], $$
$$ \left[ \begin{array}{cccc} 1 & 2 & 0 & 1 \\ 0 & 0 & 1 & 2 \end{array} \right],  
\left[ \begin{array}{cccc} 1 & 0 & 2 & 0 \\ 0 & 1 & 2 & 2 \end{array} \right], \left[ \begin{array}{cccc} 1 & 0 & 1 & 1 \\ 0 & 1 & 0 & 0 \end{array} \right], 
\left[ \begin{array}{cccc} 1 & 0 & 1 & 0 \\ 0 & 1 & 0 & 1 \end{array} \right], \left[ \begin{array}{cccc} 
1 & 0 & 0 & 1 \\ 0 & 1 & 1 & 0 \end{array} \right], $$
$$ \left[ \begin{array}{cccc} 1 & 0 & 0 & 2 \\ 0 & 0 & 1 & 1 \end{array} \right],  
\left[ \begin{array}{cccc} 0 & 1 & 0 & 2 \\ 0 & 0 & 1 & 1 \end{array} \right], \left[ \begin{array}{cccc} 
1 & 0 & 2 & 0 \\ 0 & 1 & 0 & 1 \end{array} \right], 
\left[ \begin{array}{cccc} 1 & 0 & 0 & 1 \\ 0 & 1 & 2 & 0 \end{array} \right], 
\left[ \begin{array}{cccc} 
1 & 0 & 1 & 1 \\ 0 & 1 & 2 & 0 \end{array} \right], $$
$$ \left[ \begin{array}{cccc} 1 & 0 & 2 & 0 \\ 0 & 1 & 1 & 1 \end{array} \right],  
\left[ \begin{array}{cccc} 1 & 0 & 0 & 0 \\ 0 & 0 & 1 & 1 \end{array} \right], 
\left[ \begin{array}{cccc} 
1 & 0 & 1 & 0 \\ 0 & 0 & 0 & 1 \end{array} \right], 
\left[ \begin{array}{cccc} 1 & 0 & 1 & 0 \\ 0 & 1 & 2 & 1 \end{array} \right], 
\left[ \begin{array}{cccc} 
1 & 0 & 2 & 1 \\ 0 & 1 & 1 & 0 \end{array} \right], $$
$$ \left[ \begin{array}{cccc} 1 & 0 & 2 & 1 \\ 0 & 1 & 2 & 0\end{array} \right],  
\left[ \begin{array}{cccc} 1 & 0 & 2 & 0 \\ 0 & 1 & 2 & 1 \end{array} \right], 
\left[ \begin{array}{cccc} 
1 & 0 & 2 & 2 \\ 0 & 1 & 1 & 2 \end{array} \right], 
\left[ \begin{array}{cccc} 1 & 0 & 0 & 0 \\ 0 & 1 & 0 & 0 \end{array} \right], 
\left[ \begin{array}{cccc} 
0 & 1 & 0 & 0 \\ 0 & 0 & 0 & 1 \end{array} \right]. $$
The ternary code $\C_b'$ with generator matrix $G_a$ is cyclic so the $H_{23}$-code $a\,\C_a + b\,\C_b'$ is a cyclic code by Corollary \ref{cyclic}.

There is a unique binary self-dual code $\C_a$ of length $6$ with generator matrix \[\left[ \begin{array}{cccccc} 1 & 0 & 0 & 0 & 0 & 1 \\ 0 & 1 & 0 & 0 & 1 & 0 \\ 0 & 0 & 1 & 1 & 0 & 0 \end{array} \right].\] Up to equivalence, we find $1032$ inequivalent QSD codes over $H_{23}$ of length $6$ . We found more QSD codes over $H_{23}$ for lengths $n=2,4,6$ than \cite{AAABBSS}.

\subsection{Self-orthogonal codes over $H_{32}$}
If $(k_a,k_b)=(n,0)$, then we have the self-orthogonal code $\C=a\mathbb{F}_2^n$.
If $(k_a,k_b)=(0,k_b)$ or $(k_a,k_b)=(n,k_b)$, then the number of inequivalent self-orthogonal codes over $H_{32}$
for a given length $n$ is equal to the number of permutation inequivalent self-orthogonal ternary codes of dimension $k_b$.
If $(k_a,k_b)=(k_a,0)$, then the number of inequivalent self-orthogonal codes over $H_{32}$
for a given length $n$ is equal to the number of inequivalent binary codes of dimension $k_a$.
For the other cases,  please refer to Table 4.



\begin{table}[ht]\label{tab:classifyH32}
\caption{Number of nonzero inequivalent self-orthogonal $H_{32}$-codes of lengths $n$ up to $7$}
\begin{center}

{\footnotesize 
\begin{tabular}{cccccc|cccccc}
\hline 
$n$ & $(k_{a},k_{b})$ & $|L_a|$ &  $|L_b|$ & $\#$Codes & Remark & $n$ & $(k_{a},k_{b})$ & $|L_a|$ &  $|L_b|$ & $\#$Codes & Remark \tabularnewline
\hline 
$3$ & $(1,1)$ & $3$ &  $2$ & 8 &   & $6$ & $(2,2)$ & $16$ &  $4$ & $469$ &  \tabularnewline
$3$ & $(2,1)$ & $3$ &  $2$ & 8 & & $6$ & $(3,2)$ & $22$ &  $4$ & $866$ &  \tabularnewline
\cline{1-6} \cline{2-6} \cline{3-6} \cline{4-6}  \cline{5-6} \cline{6-6} 
$4$ & $(1,1)$ & $4$ &  $2$ & 18 & & $6$ & $(4,2)$ & $16$ &  $4$ & $469$ &  \tabularnewline
$4$ & $(2,1)$ & $6$ &  $2$ & 35 & &  $6$ & $(5,2)$ & $6$ &  $4$ & $75$ &  \tabularnewline
\cline{7-12} \cline{8-12} \cline{9-12} \cline{10-12}  \cline{11-12} \cline{12-12} 
$4$ & $(3,1)$ & $4$ &  $2$ & 18 & & $7$ & $(1,1)$ & $7$ &  $6$ & $132$ &   \tabularnewline
$4$ & $(1,2)$ & $4$ &  $1$ & 7 & & $7$ & $(2,1)$ & $23$ &  $6$ & $963$ &  \tabularnewline
$4$ & $(2,2)$ & $6$ &  $1$ & 13 & QSD & $7$ & $(3,1)$ & $43$ &  $6$ & $2623$ &  \tabularnewline

$4$ & $(3,2)$ & $4$ &  $1$ & 7 &  & $7$ & $(4,1)$ & $43$ &  $6$ & $2623$ &  \tabularnewline
\cline{1-6} \cline{2-6} \cline{3-6} \cline{4-6}  \cline{5-6} \cline{6-6} 
$5$ & $(1,1)$ & $5$ &  $2$ & 28 &  & $7$ & $(5,1)$ & $23$ &  $6$ & $963$ &  \tabularnewline 

$5$ & $(2,1)$ & $10$ &  $2$ & 99 &  & $7$ & $(6,1)$ & $7$ &  $6$ & $132$ &  \tabularnewline
$5$ & $(3,1)$ & $10$ &  $2$ & 99 &  & $7$ & $(1,2)$ & $7$ &  $10$ & $354$ &  \tabularnewline
$5$ & $(4,1)$ & $5$ &  $2$ & 28  &  & $7$ & $(2,2)$ & $23$ &  $10$ & $3999$ &  \tabularnewline
$5$ & $(1,2)$ & $5$ &  $1$ & $15$  &  & $7$ & $(3,2)$ & $43$ &  $10$ & $13675$ & \tabularnewline
$5$ & $(2,2)$ & $10$ &  $1$ & $57$  &  & $7$ & $(4,2)$ & $43$ &  $10$ & $13675$ &  \tabularnewline
$5$ & $(3,2)$ & $10$ &  $1$ & $57$ &  & $7$ & $(5,2)$ & $23$ &  $10$ & $3999$ &  \tabularnewline
$5$ & $(4,2)$ & $5$ &  $1$ & $15$   &  & $7$ & $(6,2)$ & $7$ &  $10$ & $354$ &  \tabularnewline
\cline{1-6} \cline{2-6} \cline{3-6} \cline{4-6}  \cline{5-6} \cline{6-6} 
$6$ & $(1,1)$ & $6$ &  $6$ & $78$  &  & $7$ & $(1,3)$ & $7$ &  $2$ & $78$ &  \tabularnewline
$6$ & $(2,1)$ & $16$ &  $6$ & $360$ &  & $7$ & $(2,3)$ & $23$ &  $2$ & $939$ &  \tabularnewline
$6$ & $(3,1)$ & $22$ &  $6$ & $603$ &  & $7$ & $(3,3)$ & $43$ &  $2$ & $3405$ &  \tabularnewline
$6$ & $(4,1)$ & $16$ &  $6$ & $360$ &  & $7$ & $(4,3)$ & $43$ &  $2$ & $3405$ &  \tabularnewline
$6$ & $(5,1)$ &  $6$ &  $6$ & $78$ &  & $7$ & $(5,3)$ & $23$ &  $2$ & $939$ &  \tabularnewline
$6$ & $(1,2)$ & $6$ &  $4$ & $75$  &  & $7$ & $(6,3)$ & $7$ &  $2$ & $78$  &  \tabularnewline

\hline
\end{tabular}}
\end{center}
\end{table}
There is a unique ternary self-dual code up to permutation equivalence of length $4$ with generator matrix \[ \C_b=\left[ \begin{array}{cccc} 1 & 0 & 1 & 1 \\ 0 & 1 & 1 & 2 \end{array} \right].\] 
Up to equivalence, we find $13$ QSD codes of length $4$ over $H_{32}$. This result agrees with  \cite{AAABBSS}
and by Corollary \ref{cyclic}, none of these codes is a cyclic code. The pairs $(\C_a,\sigma_i(\C_b))$ that corresponds to the code $a\,\C_a + b\,\sigma_i(\C_b))$ are listed in Table 5.

\begin{table}[ht]\label{tab:QSDH32n4}
\caption{The 13 inequivalent self-orthogonal $H_{32}$-codes of length $n=4$}
\begin{center}
{\small \begin{tabular}{|c|c|c|c|}
\hline
$\C_a$ & $\sigma_i(\C_b)$ & $\C_a$ & $\sigma_i(\C_b)$ \\ 
\hline \hline
$\displaystyle \left[ \begin{array}{cccc} 1 & 1 & 0 & 1 \\ 0 & 0 & 1 & 0 \end{array} \right]$ & 
$\displaystyle \left[ \begin{array}{cccc} 1 & 0 & 1 & 1 \\ 0 & 1 & 1 & 2 \end{array} \right]$ &
$\displaystyle \left[ \begin{array}{cccc} 1 & 1 & 0 & 1 \\ 0 & 0 & 1 & 0 \end{array} \right]$ & 
$\displaystyle \left[ \begin{array}{cccc} 1 & 0 & 2 & 2 \\ 0 & 1 & 1 & 2 \end{array} \right]$   \\
\hline
$\displaystyle \left[ \begin{array}{cccc} 0 & 1 & 0 & 1 \\ 0 & 0 & 1 & 1 \end{array} \right]$ & 
$\displaystyle \left[ \begin{array}{cccc} 1 & 0 & 1 & 1 \\ 0 & 1 & 1 & 2 \end{array} \right]$ &
$\displaystyle \left[ \begin{array}{cccc} 0 & 1 & 0 & 1 \\ 0 & 0 & 1 & 1 \end{array} \right]$ & 
$\displaystyle \left[ \begin{array}{cccc} 1 & 0 & 1 & 2 \\ 0 & 1 & 1 & 1 \end{array} \right]$   \\
\hline
$\displaystyle \left[ \begin{array}{cccc} 1 & 1 & 0 & 0 \\ 0 & 0 & 1 & 1 \end{array} \right]$ & 
$\displaystyle \left[ \begin{array}{cccc} 1 & 0 & 1 & 1 \\ 0 & 1 & 1 & 2 \end{array} \right]$ &
$\displaystyle \left[ \begin{array}{cccc} 1 & 0 & 0 & 1 \\ 0 & 0 & 1 & 0 \end{array} \right]$ & 
$\displaystyle \left[ \begin{array}{cccc} 1 & 0 & 1 & 1 \\ 0 & 1 & 1 & 2 \end{array} \right]$   \\
\hline
$\displaystyle \left[ \begin{array}{cccc} 1 & 0 & 0 & 1 \\ 0 & 0 & 1 & 0 \end{array} \right]$ & 
$\displaystyle \left[ \begin{array}{cccc} 1 & 0 & 2 & 2 \\ 0 & 1 & 1 & 2 \end{array} \right]$ &
$\displaystyle \left[ \begin{array}{cccc} 1 & 0 & 0 & 1 \\ 0 & 0 & 1 & 0 \end{array} \right]$ & 
$\displaystyle \left[ \begin{array}{cccc} 1 & 0 & 1 & 2 \\ 0 & 1 & 1 & 1 \end{array} \right]$   \\
\hline
$\displaystyle \left[ \begin{array}{cccc} 1 & 0 & 0 & 1 \\ 0 & 0 & 1 & 0 \end{array} \right]$ & 
$\displaystyle \left[ \begin{array}{cccc} 1 & 0 & 2 & 1 \\ 0 & 1 & 2 & 2 \end{array} \right]$ &
$\displaystyle \left[ \begin{array}{cccc} 1 & 0 & 0 & 0 \\ 0 & 0 & 0 & 1 \end{array} \right]$ & 
$\displaystyle \left[ \begin{array}{cccc} 1 & 0 & 1 & 1 \\ 0 & 1 & 1 & 2 \end{array} \right]$   \\
\hline
$\displaystyle \left[ \begin{array}{cccc} 1 & 0 & 0 & 0 \\ 0 & 0 & 0 & 1 \end{array} \right]$ & 
$\displaystyle \left[ \begin{array}{cccc} 1 & 0 & 1 & 2 \\ 0 & 1 & 1 & 1 \end{array} \right]$ &
$\displaystyle \left[ \begin{array}{cccc} 1 & 0 & 1 & 1 \\ 0 & 1 & 1 & 1 \end{array} \right]$ & 
$\displaystyle \left[ \begin{array}{cccc} 1 & 0 & 1 & 1 \\ 0 & 1 & 1 & 2 \end{array} \right]$   \\
\hline
$\displaystyle \left[ \begin{array}{cccc} 1 & 0 & 1 & 1 \\ 0 & 1 & 1 & 1 \end{array} \right]$ & 
$\displaystyle \left[ \begin{array}{cccc} 1 & 0 & 2 & 2 \\ 0 & 1 & 2 & 1 \end{array} \right]$ & & \\
\hline
\end{tabular} }

\end{center}
\end{table}
Generator matrices of the pairs $(\C_a,\sigma_i(\C_b))$ that corresponds to $H_z$-codes in Table 3 and Table 4 
may be obtained by the interested reader from the authors.

\end{document}